\documentclass{elsart}

 \usepackage{graphicx}

\def\d{{\rm d}}

\begin{document}

\begin{frontmatter}



\title
{Equilibrium traffic flow of a mixture of 
cars with different properties}

\author{Anton \v Surda}

\address{Institute of Physics, SAS, 815 00 Bratislava, Slovakia}

\ead{anton.surda@savba.sk}

\begin{abstract}
Statistical mechanics of a disordered system of cars on a single-lane road is 
developed. Behaviour of cars is defined by conditional probability of car 
velocity depending on the distance and velocity of the car ahead. A system 
consisting of different cars is modelled by a system of two types of cars 
differing in maximal velocity or efficiency of brakes. Starting from 
conditional probabilities and using principle of maximum entropy, probability 
densities of car velocities and headways are calculated.  It is shown that the 
first-order phase transition between free flow and congested traffic may be 
driven by number of fast cars in a system of slow cars, and, as a rule, 
admixture of cars of superior qualities does not increase but decreases the 
total flow. In the system of cars with poor brakes platoons of cars of the 
same velocity are formed. They  are dissolved by a small addition of cars with 
good brakes.
Application of principle of maximum entropy was justified by comparing the 
results with steady state properties of an equivalent kinetic model.
\end{abstract}

\end{frontmatter}

\section{Introduction}
\label{intr} 

The behaviour of cars on a single-lane road without possibility of overtaking
is given by  reaction of each driver on the velocity and distance (headway)  
of the car ahead. The drivers cannot influence the headway directly, they 
control only the velocity which is limited by the maximum construction 
velocity of the cars and necessity to avoid an accident with the car ahead if 
it is slowing 
down or stops. The driver does not react on the traffic situation always in 
the same way, so that the velocity of his car would be distributed around some 
optimum value with a probability that can be found experimentally or modelled, 
starting from simple considerations. At the beginning the model is formulated 
as continuous, with arbitrary coordinates and velocities from given intervals, 
nevertheless, the numerical calculations are performed on a one-dimensional 
lattice where the car acquires only discrete velocities. Then the model 
resembles discrete particle hopping models \cite{l:1,l:1a,l:1b}, but the 
configuration space of 
particles is enlarged from a coordinate to the coordinate and velocity space, 
and instead of hopping probabilities,  particle velocities are introduced. The 
model is applied to a steady state traffic, i.e., to a system of cars of 
constant global density observed for a long period of time.  In the real 
traffic the cars are not identical. This fact and its influence on 
the distribution of velocities and headways of the cars is exemplified by 
investigation of a random mixture of two kinds of cars.  

Our model of traffic and the method of its solution is described in Section~2.
In Section~3, the results in form of 3D plots of probability density of 
car velocity and headway averaged over a small groups of cars are presented.

\section{Model of traffic flow} 
\label{model}

The behaviour of the $i$-th car is further described by a conditional 
probability density 
$p_i(v_i|v_{i-1},x_{i-1,i}) $ of its velocity $v_i$  depending on the distance 
(headway)  and velocity of the preceding car \cite{l:2}. The probability  
density  $p_i$ can be 
found experimentally observing a couple of cars for a long enough period. 
The knowledge of the probability gives us full information of driver's 
response to traffic situations, and no assumptions on reaction times,  
breaking abilities, engine power, etc., are necessary. 
The driver can directly influence only the velocity of the car, not
its relative position with respect to its neighbours. 
There is no {\it a priori} 
information about the distances between the cars, and they  will be obtained 
as a result of calculations. 
This fact is taken into account by maximization of Shannon's 
information entropy which is identical with Gibbs entropy in statistical 
mechanics.  
The probability
distribution of distances is obtained from it together with the requirement 
that 
all the cars have the same mean velocity as they cannot overtake each other.

The states of the system are characterized by the values of velocities and 
coordinates of the vehicles, and our task is to find the probability of each 
of such states, ${\cal P}(v_1,x_{1,2},v_2,\dots,x_{N-1,N},v_N)$. (The cars are 
numbered from the left to the right and are moving to the left.)  As each car 
interacts 
only with the preceding car, fixing the velocity of one car divides the system 
into two independent parts, i.e., 
\begin{eqnarray} 
\label{eq1} 
& & {\cal P}(v_1,x_{1,2},v_2,\dots,v_i={\rm const},\dots, 
x_{N-1,N},v_N)={}\qquad\nonumber\\ 
& & {}={\cal P}_1(v_1,x_{1,2},v_2,\dots,x_{i-1,i})
{\cal P}_2(x_{i,i+1}, v_{i+1},\dots, x_{N-1,N},v_N)\nonumber
\end{eqnarray} 
where ${\cal P}_2$ depends on the value of $v_i$ and is, in fact, a 
conditional probability.

Applying the above consideration to each vehicle, we see that      
the probability that the velocities of cars in a group are
 $v_i$ and the distances between them $x_{i-1,i}$ is a product of conditional 
probabilities
\begin{equation}
{\cal P}(v_1,x_{1,2},v_2,\dots,x_{N-1,N},v_N)   = P_1(v_1)  \prod_{i=2}^{N} 
P_i(x_{i-1,i},v_{i}|v_{i-1}),  
\end{equation}
where 
\begin{eqnarray} 
\label{eq2} 
&&  P_i(x_{i-1,i},v_{i}|v_{i-1})= P_i(v_{i-1},x_{i-1,i},v_{i})/ 
P_{i}(v_{i-1}),\nonumber\\ 
&&  P_i(v_{i-1},x_{i-1,i},v_{i})=p_i(v_{i}|v_{i-1}, x_{i-1,i}) 
P_{i}(v_{i-1},x_{i-1,i}),\\ 
&&   P_{i}(v_{i-1})=\int \d{x_{i-1,i}}\,P_{i}(v_{i-1},x_{i-1,i}) 
={}\nonumber\\ 
& ={}& P_{i-1}(v_{i-1})=\int \d{x_{i-2,i-1}}\, P_{i-1}(x_{i-2,i-1},v_{i-1})
.\nonumber  
\end{eqnarray}

Substituting for $P_i(x_{i-1,i},v_{i}|v_{i-1})$ from (2)  into 
(1)  we obtain
\begin{eqnarray}
& & {\cal P}(v_1,x_{1,2},v_2,\dots,x_{N-1,N},v_N)   = {}\nonumber\\
& ={}& P_1(v_1) \prod_{i=2}^{N}  
p_i(v_{i}|v_{i-1}, x_{i-1,i}) 
{P_{i}(v_{i-1},x_{i-1,i})\over P_{i}(v_{i-1})}.   
\end{eqnarray}

The probability $\cal P$ has a relatively simple form as we assume that the 
driver of $i$-th car reacts only on the velocity of the $i-1$-th car and its 
distance $ x_{i-1,i}$ and not on the velocities and distances of the other 
cars. The conditional probability  density   $p_i(v_{i}|v_{i-1}, x_{i-1,i}) $
contains all the known information of the system, but it 
  cannot alone 
determine the  probability of  states of a group of cars. To find it, having 
no further information available, the 
principle of maximum entropy is used.  To calculate the probability $\cal 
P$, we have 
to maximize entropy $-\sum {\cal P} \log {\cal P}$ under conditions that the 
mean length of 
the system is $L $ and  the mean velocity of each  car is $\bar v$, 
which, in equilibrium, has to be the same for each car. 
Using the method of Lagrange multipliers, we have to find the extreme of the 
expression 
\eject

\begin{eqnarray}
& & \int\prod_i\d x_{i,i+1} \d v_{i}\,  {\cal P} \log {\cal P}+{}\nonumber\\  
& {}+{}&  \mu \sum_{i}^{}\int \d{x_{i-i,i}}\, x_{i-1,i}\int \d{v_{i-1}}\, 
P_{i}(v_{i-1},x_{i-1,i})+{}\nonumber \\
&   +{}&  
\sum_{i}\nu_{i}\int \d{v_{i}}\, v_{i} 
\int\d{v_{i-1}}\int\d{x_{i-1,i}}\,p_i(v_{i}|v_{i-1}, x_{i-1,i})
P_{i}(v_{i-1},x_{i-1,i})+{} \\  
& {}+{}& 
\sum_{i}\lambda _i\int\d{v_{i-1}} \int \d{x_{i-1,i}}\,P_{i}(v_{i-1},x_{i-1,i}) 
\nonumber  
\end{eqnarray}
where  $\mu$, $\nu_i$, $\lambda_i$ are Lagrange multipliers.

Substituting (3) into (4)  and maximizing it with respect to  
$P_{i}(v_{i-1},x_{i-1,i})$, we get
\begin{equation}
\hskip-5mm\log {P_{i}(v_{i-1},x_{i-1,i})\over P_{i}(v_{i-1})}+\mu x_{i-1,i} 
+\nu_i \bar v_{i}(v_{i-1},x_{i-1,i}) + s_{i}(v_{i-1},x_{i-1,i})+\lambda_i=0  
\end{equation}
where $\bar v_{i}(v_{i-1},x_{i-1,i})=\int\d{v_{i}}\, v_{i} 
p_i(v_{i}|v_{i-1}, x_{i-1,i})$ and
$s_{i}(v_{i-1},x_{i-1,i}) =\break \int\d{v_{i}}\, 
p_{i}(v_{i}|v_{i-1},x_{i-1,i})  \cdot  
\log  p_{i}(v_{i}|v_{i-1},x 
_{i-1,i} )$ are the mean velocity and entropy of a car with fixed 
boundary conditions.
 $\int\d{x_{i-1,i}}\, P_i(v_{i-1},x_{i-1,i})= 
P_{i-1}(v_{i-1})$ is also required. Then, using (2),(3)  and (5), the 
probabilities appearing in (1) can be calculated from the following formula  
\begin{eqnarray}
&&   P_i(x_{i-1,i},v_{i}|v_{i-1}) = p_{i}(v_{i}|v_{i-1},x_{i-1,i}) 
P_i(x_{i-1,i}|v_{i-1}) \\
& & P_i(x_{i-1,i}|v_{i-1})=
 A(v_{i-1}) 
\exp(-\mu x_{i-1,i} -\nu_i \bar v_{i}(v_{i-1},x_{i-1,i})-s_{i}(v_{i-1},x_{i-1,i
} )),\nonumber
\end{eqnarray}
where $A(v_{i-1})$ is the normalization constant 
$$
\hskip-7mm A(v_{i-1})=\left[\int\d{x_{i-1,i}}
\exp(-\mu x_{i-1,i} -\nu_i \bar v_{i}(v_{i-1},x_{i-1,i})-s_{i}(v_{i-1},
x_{i-1,i}))  
\right] ^{-1}\!\!\!\! .
$$

Prescribing the mean velocity and density of cars, the Langrange multipliers 
$\nu_i$ and $\mu$, respectively, can be calculated  from these requirements. 
In a system of identical cars all $\nu_i$ would  be 
the same and equal to global $\nu$.  As the velocity is a nonconserved 
quantity, $\nu=0$ in  full equilibrium, where the velocities are not given, 
but  obtained from the requirement of maximum entropy. 
  Multipliers $\nu_i$  are non-zero for cars with an obstacle in front of 
them, causing reduction 
their velocity, or if the cars are not identical, and the slower cars are 
hindering the faster ones.

The statistical properties of a car are given by the probability density
that the car velocity is $v_i$ and its headway $x_{i-1, i}$ 
\begin{equation}
  P_i(x_{i-1,i},v_i) = \int\d{v_{i-1}}\,P_i(x_{i-1,i}, v_i|v_{i-1})
P_{i-1}(v_{i-1})
\end{equation}
where $P_i(x_{i-1,i}, v_i|v_{i-1})$ is calculated from (6).

The probability density $P_i(x_{i-1,i}, v_i)$ reflects the behaviour of 
a single car. Correlations between velocities and positions of different cars 
are given by  two- or many-body correlation functions. We shall not calculate 
them here. Instead of it,  the probability density $P(V,X)$ that 
the average car velocity in
a small 
group  of $m$ cars is $V=(1/m) \sum_{i=1}^m v_i$ and average headway 
$X=(1/m) \sum_{i=1}^m x_{i-1,i}$  is calculated \cite{l:4}. This quantity 
yields 
the information about the distribution of single car properties as well as 
correlations between the cars. 
The contribution of correlations to the distribution is given by the 
difference 
 between 
the probability density $P(V,X)$  and the probability that a group of 
uncorellated cars with 
probabilities $P_i(x_{i-1,i},v_i)$  have the same values of the averaged 
velocity and headway
\begin{eqnarray}
& D(V,X) & =\int\prod_{i=1}^N\d x_{i-1,i} \d v_i\,
\delta (V-\sum_i v_i/m) \delta 
(X-\sum_i x_{i-1,i}/m)\cdot{}\nonumber\\
&&  {}\cdot\bigg[\int\d{v_{0}}\prod_{i=1}^N P_i(x_{i-1,i},v_i|v_{i-1})P(v_0)  
-\prod_{i=1}^N P_i(x_{i-1,i},v_i)\bigg].
\end{eqnarray}

The conditional probability  of velocity of the $i$-th car for given headway  
and velocity of the preceding car
$p_{i}(v_{i}|v_{i-1},x_{i-1,i})$ is not acquired from experiment in this paper,
 but calculated from simple considerations \cite{l:2}, close to the ideas of 
car-following model \cite{l:3}. It is assumed to be peaked around an optimal 
velocity  $v_{i,\rm 
opt}$, which  is further chosen as 90\% of maximum safe velocity $v_{i,\rm m}$.
The maximum safe velocity is determined from the requirement that two 
neighbouring cars, which start to decelerate at the same time with their    
deceleration rates $a_{i-1}, a_i$, would stop without crash. Moreover,   
$v_{i,\rm m}$ must 
not be greater than the maximum possible velocity of the car  $v_{i,\rm max}$, 
i.e., the optimal velocity of car $i$ is
\begin{eqnarray}
&\hskip-10mm&  v_{\rm i,opt}(v_{i-1},x_{{i-1},i}) = 0.9 v_{\rm 
i,m}(v_{i-1},x_{{i-1},i}) ,\\ 
& \hskip-10mm& v_{i,\rm m}(v_{i-1},x_{{i-1},i}) 
=
\left\{ {\begin{array}{*{20}l}
-a_i\tau_i +
\sqrt{(a_i\tau_i)^2+ 2a_ix_{{i-1},i}+ v_{i-1}^2 a_i/a_{i-1}}& {\rm if}\  
v_{\rm i,m}\le v_{\rm i,max}\nonumber\\
v_{\rm i,max}\quad & {\rm if}\  v_{\rm i,m}> v_{\rm i,max}\\ 
\end{array}}\right.
\end{eqnarray} 
 where $x_{i-1,i}$ and $v_{i-1}$ are the distance (headway) and velocity of 
the car 
ahead, respectively, $a_j$ denotes the deceleration rate of car $j$ in 
time of braking, and $\tau_j$ is the reaction time of the $j$-th
driver.  

The cars in (9)  are not identical. They differ in their maximum velocities 
and 
deceleration rates. In the formula, it is assumed that the driver of a car can 
distinguish the  type of the preceding car and knows its maximum 
deceleration rate. In the opposite case the driver $i$ should expect that  
$a_{i-1}=\max a_j$ for all $j$.

The equations of the above-described model are further solved numerically. 
For numerical computations, it is more convenient to treat the model on 
a discrete lattice with velocities acquiring only integer values. This 
approach was also used in kinetic models of traffic flow  \cite{l:1}.  
Now, our model may be considered as a generalized version of the ASEP model 
\cite{l:5}. It was shown that our equilibrium model describes well the steady 
states of the ASEP model. The space dependence of density and pair correlation 
functions is very close to the exact solutions \cite{l:2}.

In numerical calculations, the cars are presented by dimensionless 
points moving on a discrete 
one-dimen\-sio\-nal lattice and are characterized by 2 quantities:  discrete 
integer velocity $v_i$ in the interval $\langle0, v_{i,\rm max}\rangle$ and a 
discrete coordinate (site number)   $x_{i}=1,2,\dots$ (further, only their 
differences $x_{i-1,i}$ will be considered). 
 $v_{i,\rm max}$ is the maximum  velocity given by the construction of the car
(the nearest integer value to $v_{\rm opt}$ from (9) is 
taken for the optimal velocity).

As the road is assumed to be homogenous, 
in the calculation no absolute 
coordinates are used, only the distances between the cars.  
 Nevertheless, the 
velocities are absolute, measured with respect to the road.

If a homogeneous  system is in a steady state for a long time, the small 
groups of $m$ cars may be chosen at arbitrary places and arbitrary moments. In 
practice, we would observe   small parts of the road (e.g. by TV cameras)  for 
the time the density of the whole system remains constant.

The way of driving of a driver is characterized by 
 distribution of probabilities of car velocities around the optimal velocity.
Here we use an extremely simple distribution, in which the probability of 
the optimal velocity is $p_0$, the probabilities of the velocities $ v_{\rm 
opt}\pm 1$ are $p_1$, while the probability of the car to have any other 
permitted velocity is $p_2$. The values of the probabilities for velocities 
higher than the maximum safe velocity are equal to 0. The sum of all 
probabilities for each 
car is equal to 1. The parameters $p_0, p_1$, and $p_2$ are the same for 
each
car, 
and the distribution depends on the headway only by means of the value of 
optimal velocity. 
For moderate densities the continuous model is described well by our discrete 
model defined by the probabilities $p$. Nevertheless,
for very high 
densities, the discreteness of the lattice is not negligible, and  
another condition  $v_i< v_{i-1}+x_{i-1,i}$ must be  imposed. 

Our approach enables us to treat a disordered system with each car having 
different properties. To exemplify the effect of disorder, we shall 
investigate a system consisting of two types of cars for two simple cases:  
 cars are  differing by i)  maximum velocity, ii)  efficiency 
of brakes.

In an equilibrium state of a large group of cars, the cars far from the 
foremost edge 
of the group have the same mean velocity. It is calculated by 
an iteration starting from the first  car of the group driving, e.g., by 
constant, optimum velocity $\bar v_1=v_1=v_{\rm opt}$, i.e., $P_1(v_1) =\delta 
_{v_1,v_{\rm opt}}$.  
Then, the velocity distribution of the second car is  
$P_2(v_2) =\sum_{v_1,x_{1,2}} P_2(x_{1,2},v_2|v_1) P_1(v_1)$ where 
$P_2(x_{1,2},v_2|v_1)$ is given by (6)  with $\nu_2=0 $ if $\bar 
v_2=\sum_{v_2} v_2P_2(v_2) < 
\bar v_1$. 
If $\bar v_2>\bar v_1$, $P(v_2) $ is recalculated with  $\nu_2>0$ slowing 
down  car 2 to the velocity of the preceding car (1), i.e., 
$\bar v_2=v_1$. This $P(v_2) $ is  used in the next 
step.  
 The same procedure is applied consequently to
all other couples of cars $i-1,i$ in the group.  
 After large enough number of steps the mean car velocity converges to a 
constant value 
$\bar v_i=\bar v$.
 The density of cars is controlled by the Lagrange multiplier (potential)  
$\mu$ which is the only free parameter of the theory (if the conditional 
probabilities $p_i$ are known from the above considerations or an experiment 
with two cars), i.e., also $\bar v$ 
is a function of $\mu$. The Lagrange multipliers $\nu_i$ are determined from 
the requirement $\bar v_i=\bar v$.
For a system consisting of identical cars, all 
$\nu_i=0$ if the velocity of the first car is large enough. In a mixed system 
$\nu_i$'s of the fast cars are positive to slow 
them down  to the mean velocity of the cars in front of them. If there are no 
obstacles slowing down the velocity of the group, at  least some multipliers 
$\nu_i$ of the slow cars are equal to zero. In the further considerations,  
the mean car velocity is not confined, and  its natural 
value 
assuming free road without any velocity limitations, except those due to 
construction of cars, is calculated.

After reaching the steady state, probability distribution $P(V,X)$  of average 
 velocity $V$ of a car 
and average headway $X$ in a small group is calculated. It corresponds to an 
experimental observation of a constant density traffic with a TV camera 
displaying only a small part of the road. Moreover, for disordered systems 
with two 
kinds of cars, the probability distribution is averaged over a large number of 
randomly selected small groups.

To justify the use of the principle of maximum entropy   in the equilibrium 
state of our system 
of cars, the results were compared with the results obtained from numerical 
simulations of steady state of an equivalent stochastic kinetic model.   
In this model particles occupy sites of a discrete 1D lattice and perform 
jumps in one direction   of  length $\Delta x_i(t)$ at  discrete times $t$ 
with conditional probability
$p_i(\Delta x_i(t)|\Delta x_{i-1}(t), x_{i,i-1}(t-1)) $, 
where $x_{i,i-1}(t-1)= 
x_{i,i-1}(t-2)+ \Delta 
x_i(t-1)- \Delta x_{i-1}(t-1)$ for $i=2,\dots,N$. The function $p_i$ is 
identical to 
the conditional probability of the velocity of the $i$-th car, $p_i$, defined 
above and used further in calculation of equilibrium properties. The jumping 
probability of the first car in the group, $P_1(\Delta x_1(t)) $ is calculated 
selfconsistently as a probability of the jumping length of the cars of the 
same kind in 
the group of $N$ cars at the time $t-1$. The probability densities of 
the headway $X$ and car velocity $V$ were averaged over 
the last five cars of a large group and large enough numbers of 
time steps after reaching the steady state. The probability densities obtained 
from the steady state of the  kinetic model 
and  from the equilibrium distribution assuming the principle of maximum 
entropy were 
compared for both ordered and disordered systems and found to be 
very close to each other.

\section{Results}

The system is described by a large number of parameters. To study only 
few typical situations some of them are further  fixed: $p_1/p_0= 0.3$, 
$\tau=0$, and the number of cars in the group, $m=5$. The others:  
spread of the car velocities $p_2/p_0$, its maximum velocity 
$v_{\rm max}$,
deceleration rate $a$, potential $\mu$, and concentration ratio of the two 
types
of cars will acquire a small number of values.
 Typical traffic situations, shown in the following plots, appear at $v_{\rm 
max}=12, 20$, $\mu= 
0.07, 0.1, 0.12$, $a=0.5, 4$, and $p_2/p_0=0.33, 0.55$.

The calculations performed for systems consisting of identical  cars 
show \cite{l:2}  that they might be found in a free flow regime or a 
congested phase. For very high densities, platoons of cars of the same 
velocity are 
formed. The free flow and congested phase may coexist, and a first order phase 
transition takes place.  It is reflected in the shape of the probability 
density 
$P(V,X)$  with two peaks. The relative height of them is 
controlled by  the 
parameters of the model as well as by concentration of cars of different kind. 
 This takes place if cars with higher maximum velocity are added to a 
system of slow cars.

\begin{figure}
\begin{center}
\includegraphics*[width=11.5cm]{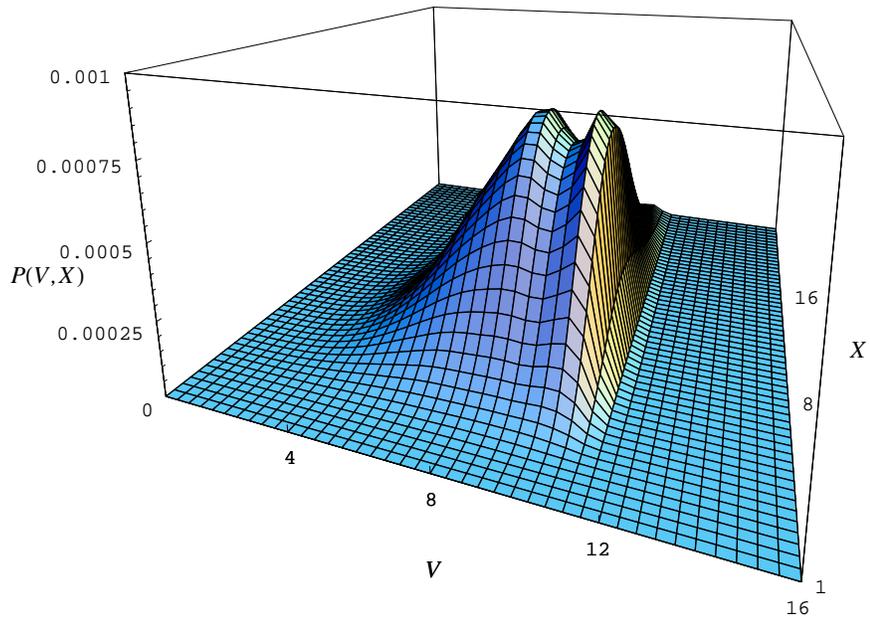}
\end{center}
\caption
{Probability density of  headway $X$ and car velocity $V$ averaged over a 
group of 5 cars 
for  $a=4$, $p_2/p_0= 0.055$, $\mu =0.1$, and $v_{\rm max}=12$.}
\label{fig:1}
\end{figure}

\begin{figure}
\begin{center}
\includegraphics*[width=11.5cm]{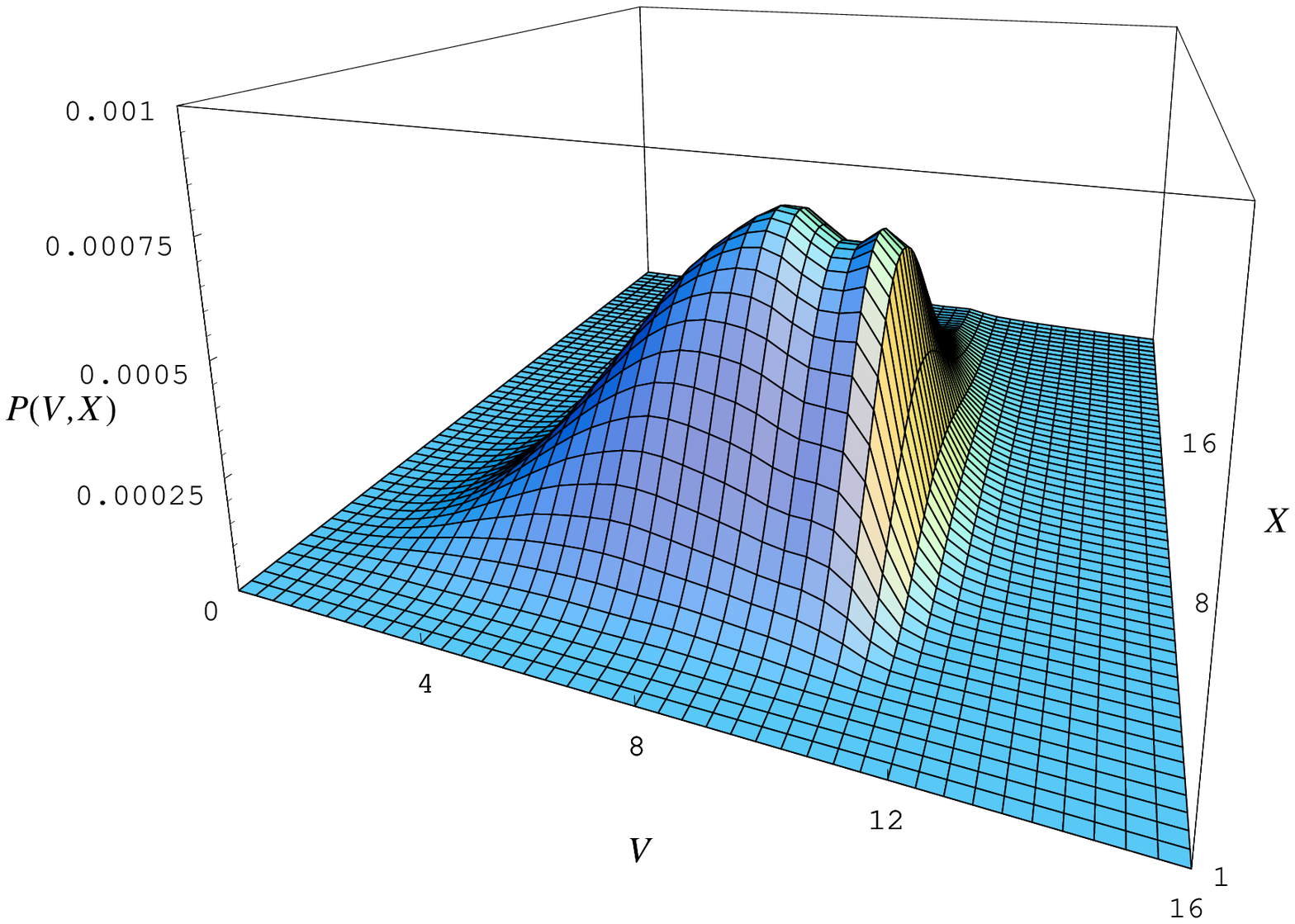}
\end{center}
\caption
{Probability density of  headway $X$ and car velocity $V$ averaged over  
groups of 5 cars 
for  $a=4$, $p_2/p_0= 0.055$, $\mu =0.1$, with 
70~\% of cars of $v_{\rm max}=12$ and
30~\% of  $v_{\rm max}=20$.
} 
\label{fig:2}
\end{figure}

\begin{figure}
\begin{center}
\includegraphics*[width=11.5cm]{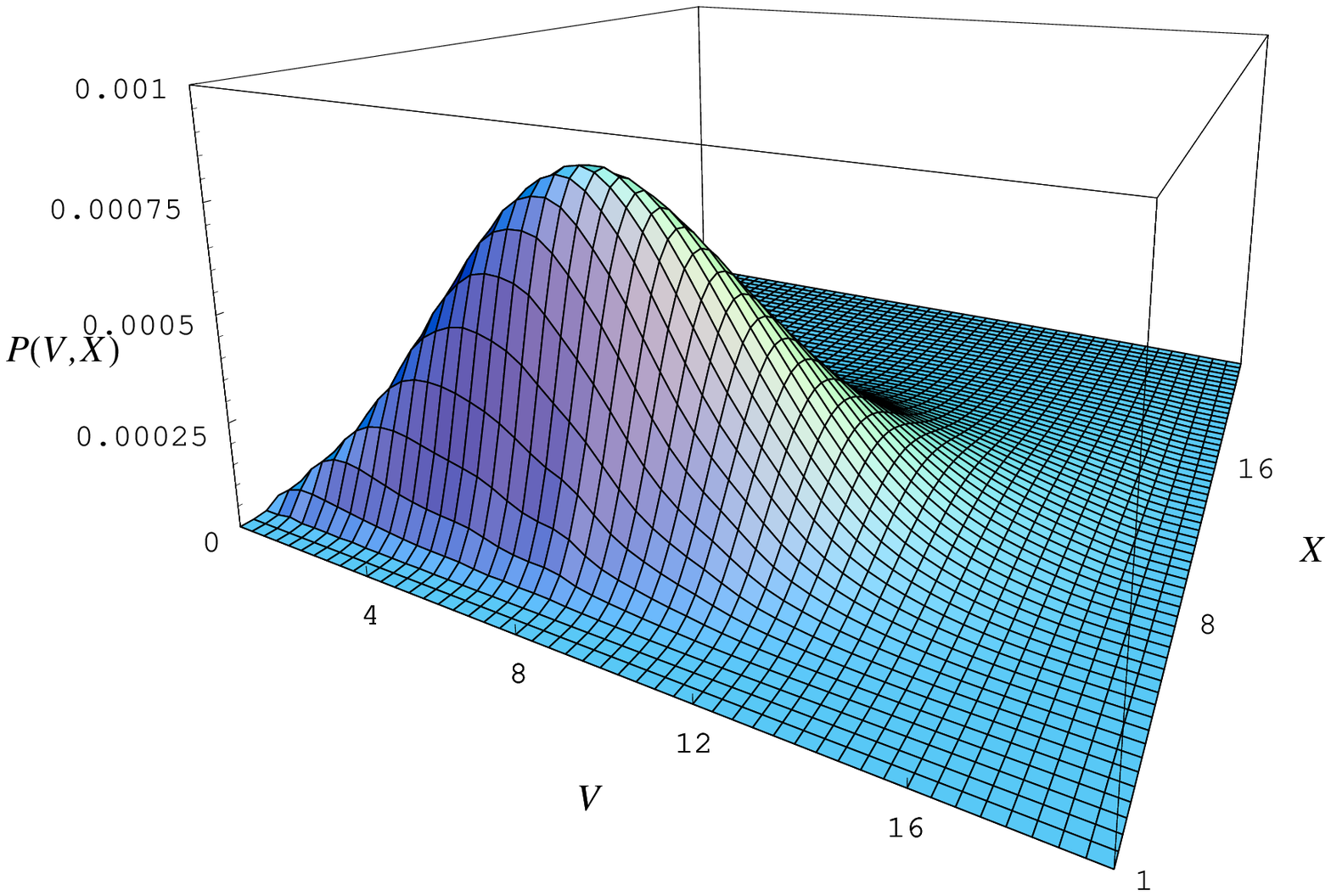}
\end{center}
\caption
{Probability density of  headway $X$ and car velocity $V$ averaged over 
groups of 5 cars 
for  $a=4$, $p_2/p_0= 0.055$, $\mu =0.1$, with 
2~\% of cars of $v_{\rm max}=12$ and
98~\% of  $v_{\rm max}=20$.
} 
\label{fig:3}
\end{figure}

\begin{figure}
\begin{center}
\includegraphics*[width=11.5cm]{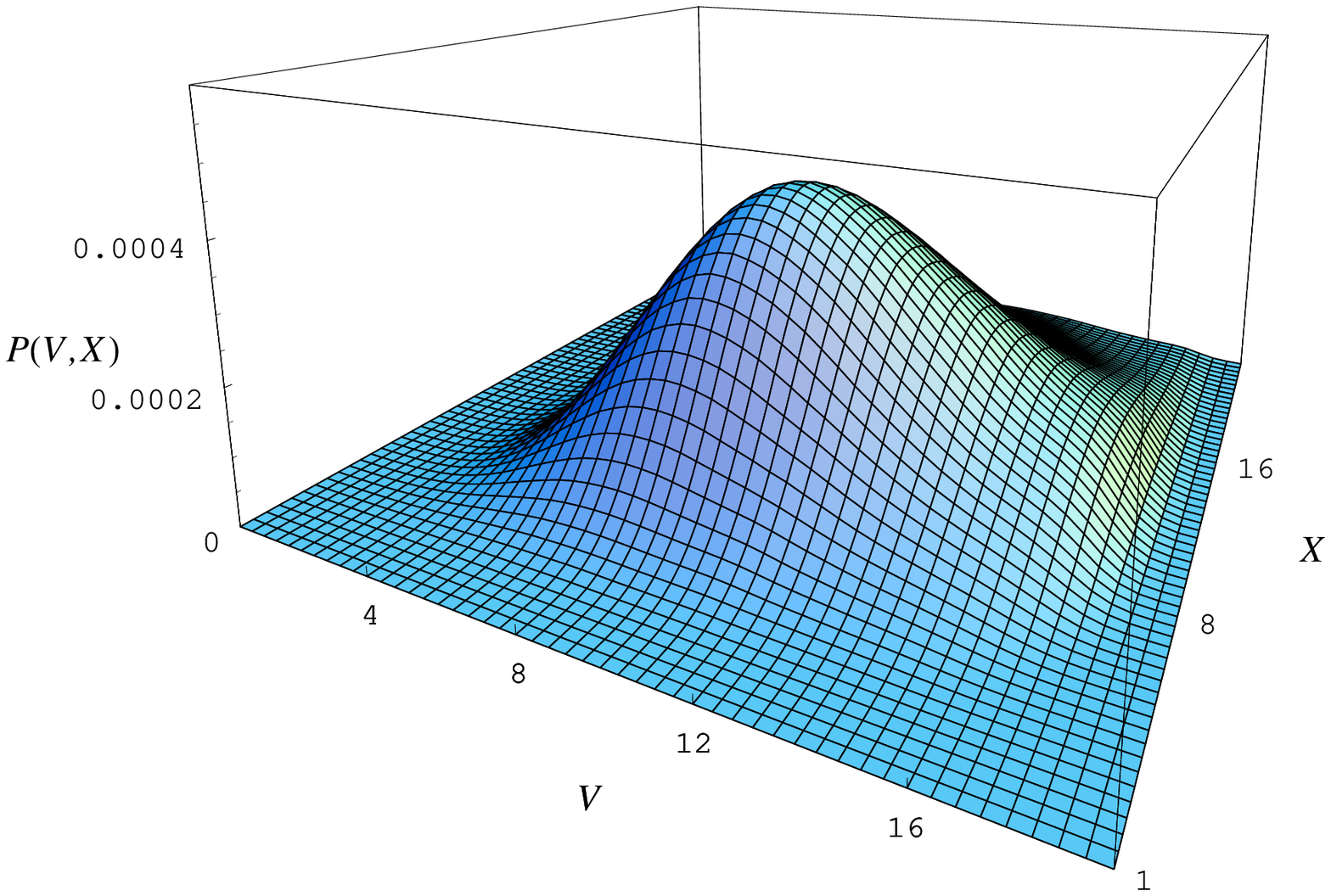}
\end{center}
\caption
{Probability density of  headway $X$ and car velocity $V$ averaged over a 
group of 5 cars 
for  $a=4$, $p_2/p_0= 0.055$, $\mu =0.1$, with 
100~\% of cars of $v_{\rm max}=20$.
} 
\label{fig:4}
\end{figure}

\begin{figure}
\begin{center}
\includegraphics*[width=11.5cm]{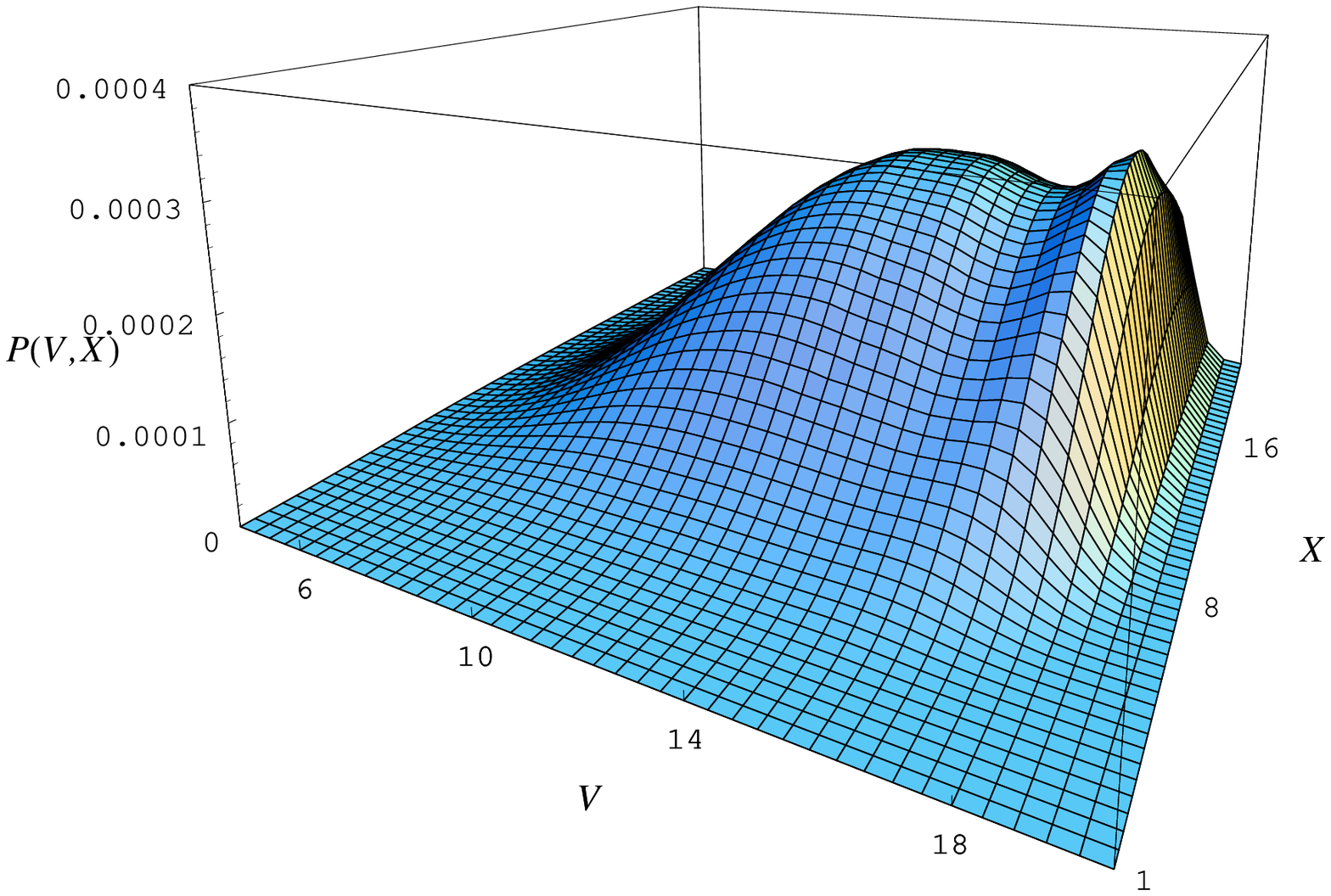}
\end{center}
\caption
{Probability density of  headway $X$ and car velocity $V$ averaged over a 
group of 5 cars
for  $a=4$, $p_2/p_0= 0.033$, $\mu =0.07$, with 
100~\% of cars of $v_{\rm max}=20$.
} 
\label{fig:5}
\end{figure}

If 
all the cars have the smaller maximum velocity, $v_{\rm max}=12$, 
$p_2/p_0=0.055$, and $\mu=0.1$,
the probability of velocity has two maxima  practically of the same height 
(Fig.~1). 
Changing these two parameters, the relative heights of the peaks change, and 
eventually one of them disappears, i.e., a first order phase transition is 
observed in the system.  This phase transition is also driven by ratio of slow 
and fast cars ($v_{\rm max}=20$)  as seen from Fig.~2
where the 
increase of concentration of 
fast cars to 30~\% causes an increase of the small velocity peak with respect 
to the high velocity one. Further addition of the fast cars makes the system 
even slower 
 and denser (Fig.~3). 
Only total disappearance of slow cars causes 
an abrupt increase of mean velocity and distances between the vehicles.
Nevertheless, the flow is not free, because of large value of the parameter 
$p_2/p_0$ the system is in the congested phase (Fig.~4),
with the most 
probable velocity much smaller than $v_{\rm opt}$.
For small values of the velocity spread 
 $p_2/p_0$, potential $\mu$, and large deceleration rate, the car flow freely, 
and the probability 
diagram consists of a peak at $v_{\rm opt}$, narrow in $V$-direction and 
wide in $X$-direction.

For the fast cars with maximum velocity $v_{\rm max}=20$,
the first order phase transition takes place at $p_2/p_0=0.033$,  
$\mu=0.07$ (Fig.~5).
Now, addition of few slow cars, in distinction 
to the previous case, does not  change 
the relative height of the two maxima, but totally destroys the probability 
diagram, and a wide peak in $V$-direction and narrow in $X$-direction is 
formed. (The probability diagram is similar to that in Fig.~2.)  Density 
is higher and velocity lower than of the system consisting only of slow cars 
($v_{\rm max}=12$).  

As the cars cannot overtake each other, the most 
serious and instantaneous impact has a small admixture of cars with inferior 
properties to a free flowing pure system.  On the other hand, addition of 
faster cars to a system of slow ones does not increase the mean velocity as 
well, but now the deterioration of the traffic flow is gradual.

If the braking abilities of the cars are poor, they form platoons of vehicles  
with the same, low velocity.  The density of cars in the platoons is not 
small, and is varying in 
a large interval. In Fig.~6,
the deceleration rate $a$ is 0.5 
instead of 
$a=4$ of previous cases. $p_2/p_1$ is 0.1, potential $\mu=0.12$, and 
the maximum velocity $v_{\rm max}=20$. The presence of  platoons is manifested 
by probability density maxima at integer values of the velocity.

\begin{figure}
\begin{center}
\includegraphics*[width=11.5cm]{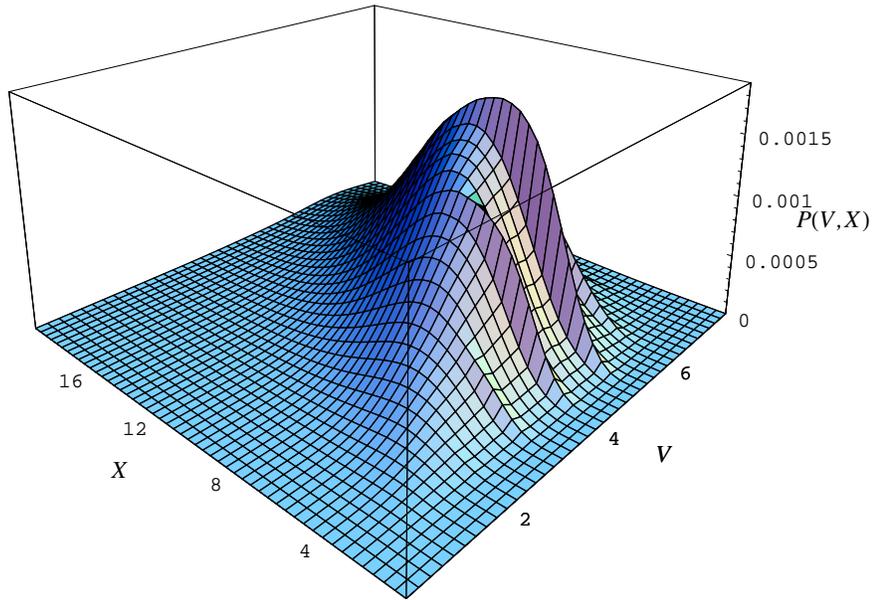}
\end{center}
\caption
{Probability density of  headway $X$ and car velocity $V$ averaged over a 
group of 5 cars 
for  $v_{\rm max}=20$, $p_2/p_0= 0.1$, $\mu =0.12$, with 
100~\% of cars of $a=0.5$.
} 
\label{fig:6}
\end{figure}

\begin{figure}
\begin{center}
\includegraphics*[width=11.5cm]{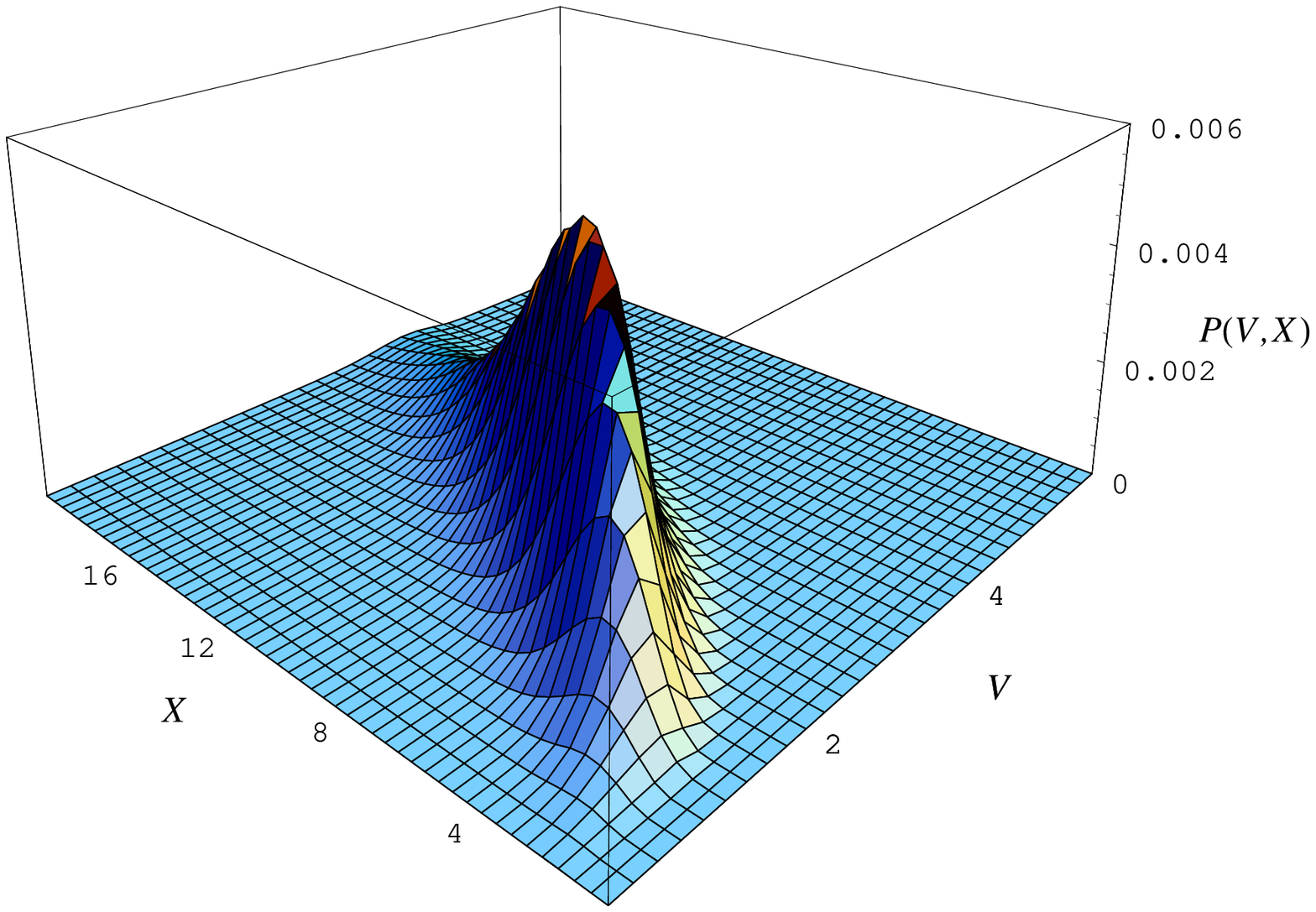}
\end{center}
\caption
{Probability density of  headway $X$ and car velocity $V$ averaged over groups 
5 {\it indistinguishable} cars
for  $v_{\rm max}=20$, $p_2/p_0= 0.1$, $\mu =0.12$, with
90~\% of cars of $a=0.5$ and
10~\% of  $a=4$.
} 
\label{fig:7}
\end{figure}

\begin{figure}
\begin{center}
\includegraphics*[width=11.5cm]{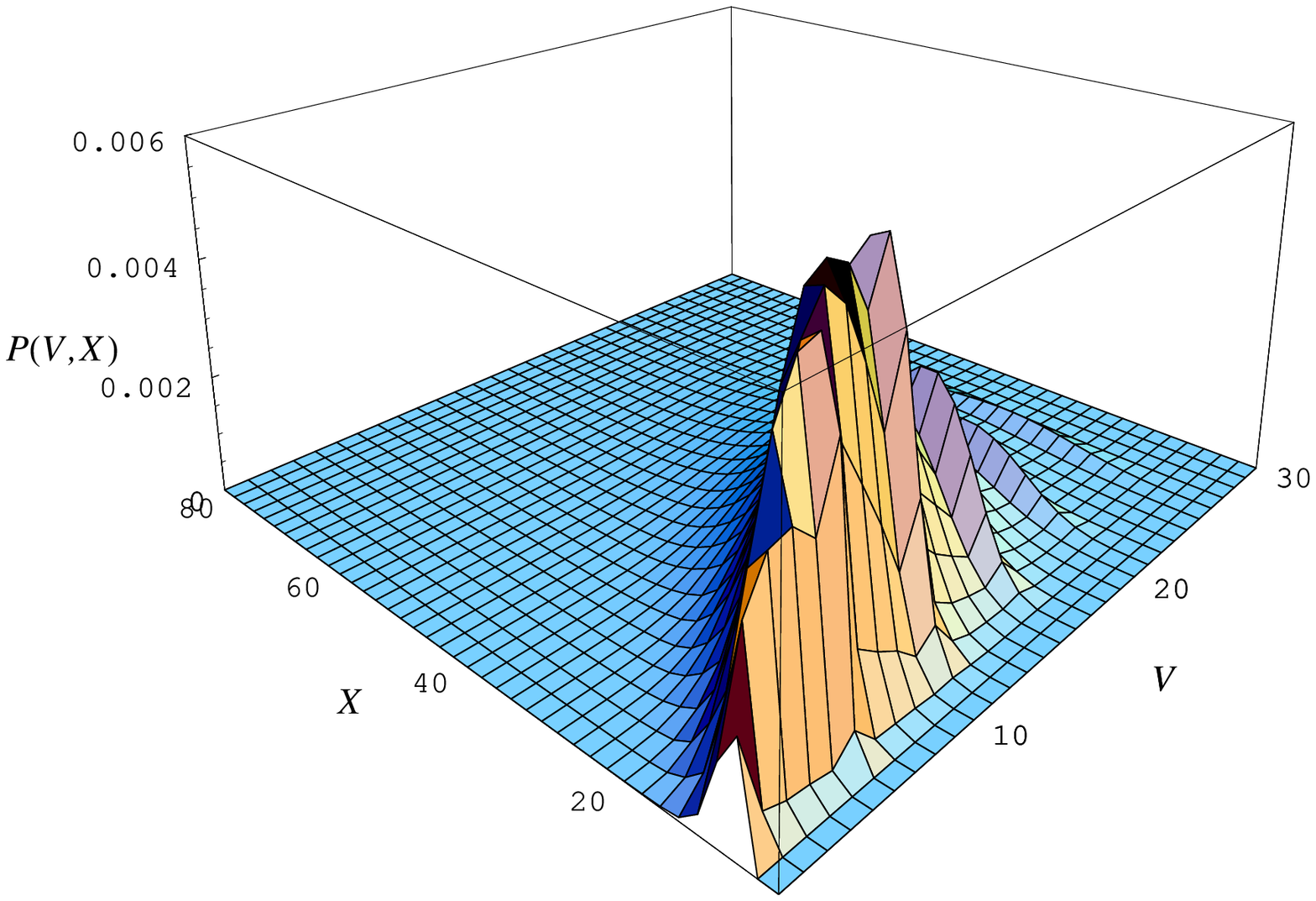}
\end{center}
\caption
{Probability density of  headway $X$ and car velocity $V$ averaged over groups 
of 5 {\it distinguishable} cars
for  $v_{\rm max}=20$, $p_2/p_0= 0.1$, $\mu =0.12$, with
90~\% of cars of $a=0.5$ and
10~\% of  $a=4$.
} 
\label{fig:8}
\end{figure}

 If few cars with good brakes $(a=4)$  are added, and the drivers are not able 
to distinguish the quality of the car ahead,  they  should 
be more careful and decrease their velocity to, approximately,  one half. 
The drivers of the cars with poor brakes  have to assume that all the cars 
have 
effective brakes and to increase the headways.  The platoons disappear, and 
the increase of the mean velocity with the mean distance between cars is slow 
(Fig.~7).

A different situation appears  
 if the quality of the brakes of the preceding car is  known to the 
driver. 
 The cars with good brakes (type 1)  may drive very closely to the cars with 
poor brakes (type 2). As  the ratio of cars with good breaks is only 10\% in 
Fig.~8,
 the cars with poor brakes drive mostly behind the cars of the same kind, 
 the drivers know that, and formation of platoons is not  hampered by 
the type 1 cars. All the cars move slowly so that also the distance 
between two cars with poor brakes is small, and the 
system is dense. Only the 
headway between type 1  and type 2 car is the same as all  the headways in the 
previous case.

For the same  parameters, but $a=4$,  the cars move practically 
freely with only a small arm of slow vehicles (Fig.~9).
This picture is 
destroyed by a small admixture (10\%)  of cars with poor brakes (Fig.~10).
These  cars should drive  cautiously 
being surrounded by cars with  good brakes which are able to stop immediately. 
They 
 form very  slow, dense platoons as at low velocity they are 
able to react effectively to slow cars in front of them. As in the front of 
 cars with ineffective brakes there are practically always cars with good 
brakes,  the possible knowledge of the type of the preceding car plays no 
role, 
and the diagram for $P(V,X) $ remains the same for both above-mentioned cases.

\begin{figure}
\begin{center}
\includegraphics*[width=11.5cm]{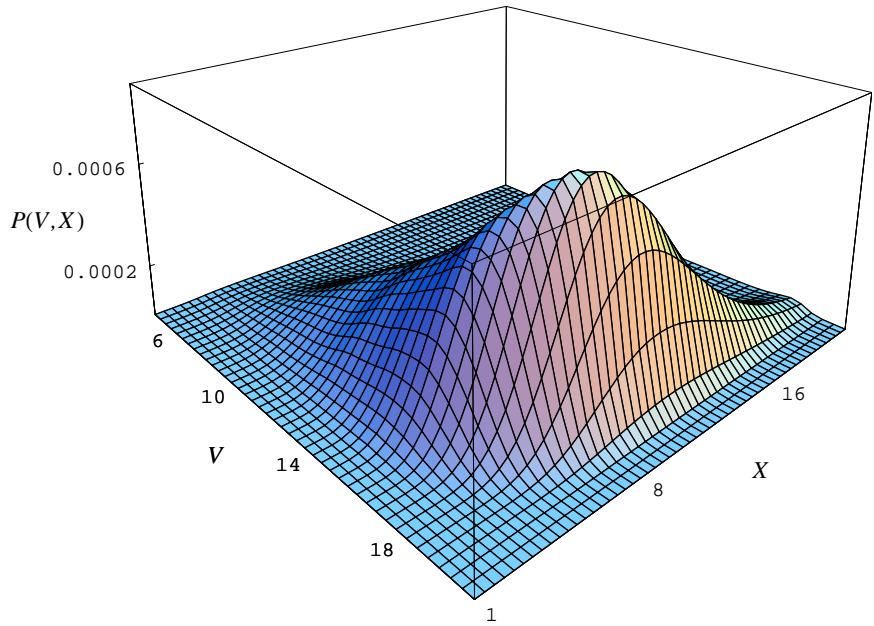}
\end{center}
\caption
{Probability density of  headway $X$ and car velocity $V$ averaged over a 
group of 5  cars 
for  $v_{\rm max}=20$, $p_2/p_0= 0.1$, $\mu =0.12$, with
100~\% of cars of $a=4$.
} 
\label{fig:9}
\end{figure}

\begin{figure}
\begin{center}
\includegraphics*[width=11.5cm]{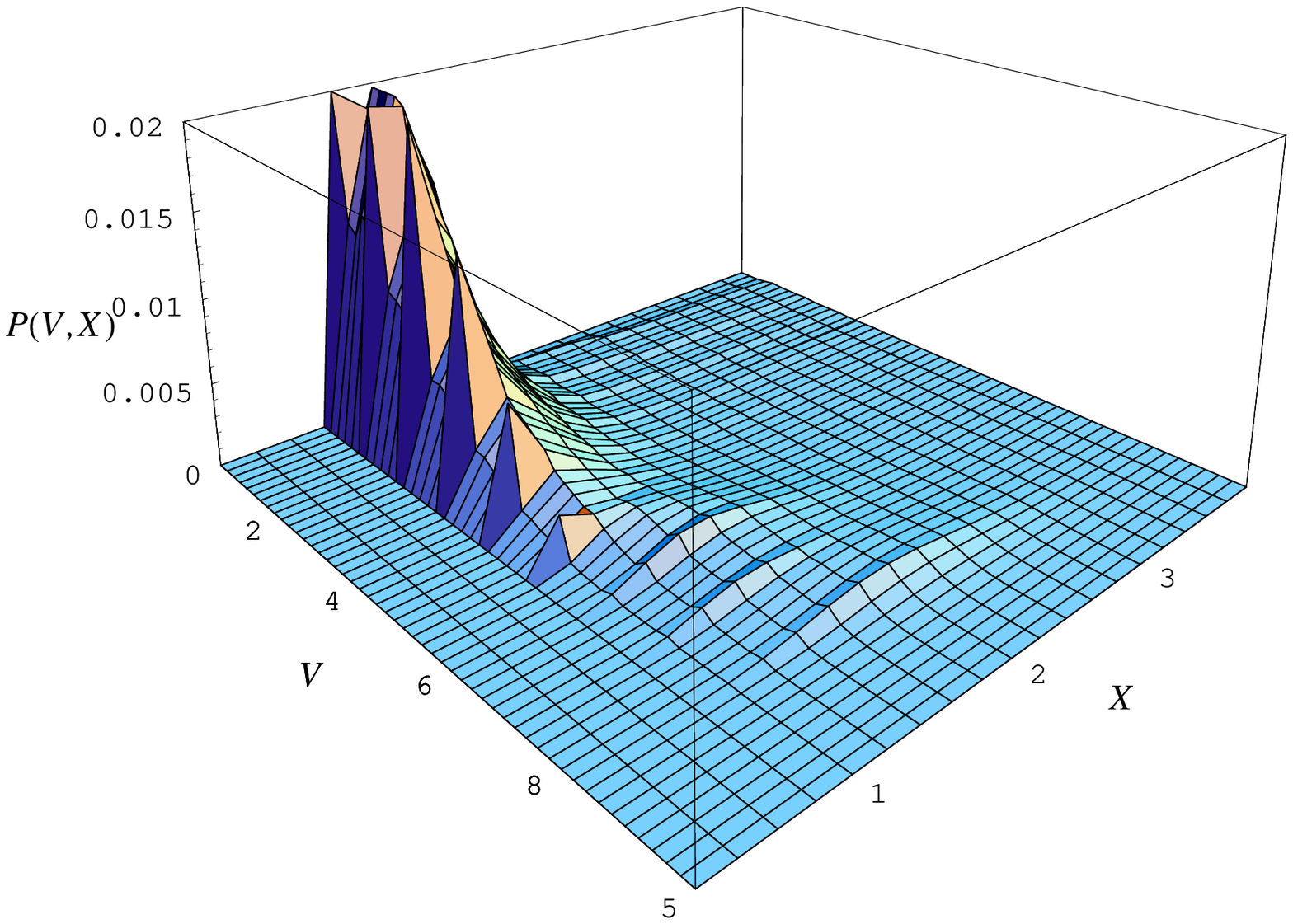}
\end{center}
\caption
{Probability density of  headway $X$ and car velocity $V$ averaged over groups 
of 5  cars 
for  $v_{\rm max}=20$, $p_2/p_0= 0.1$, $\mu =0.12$, with
10~\% of cars of $a=0.5$,
90~\% of  $a=4$.
} 
\label{fig:10}
\end{figure}

\begin{figure}
\begin{center}
\includegraphics*[width=11.5cm]{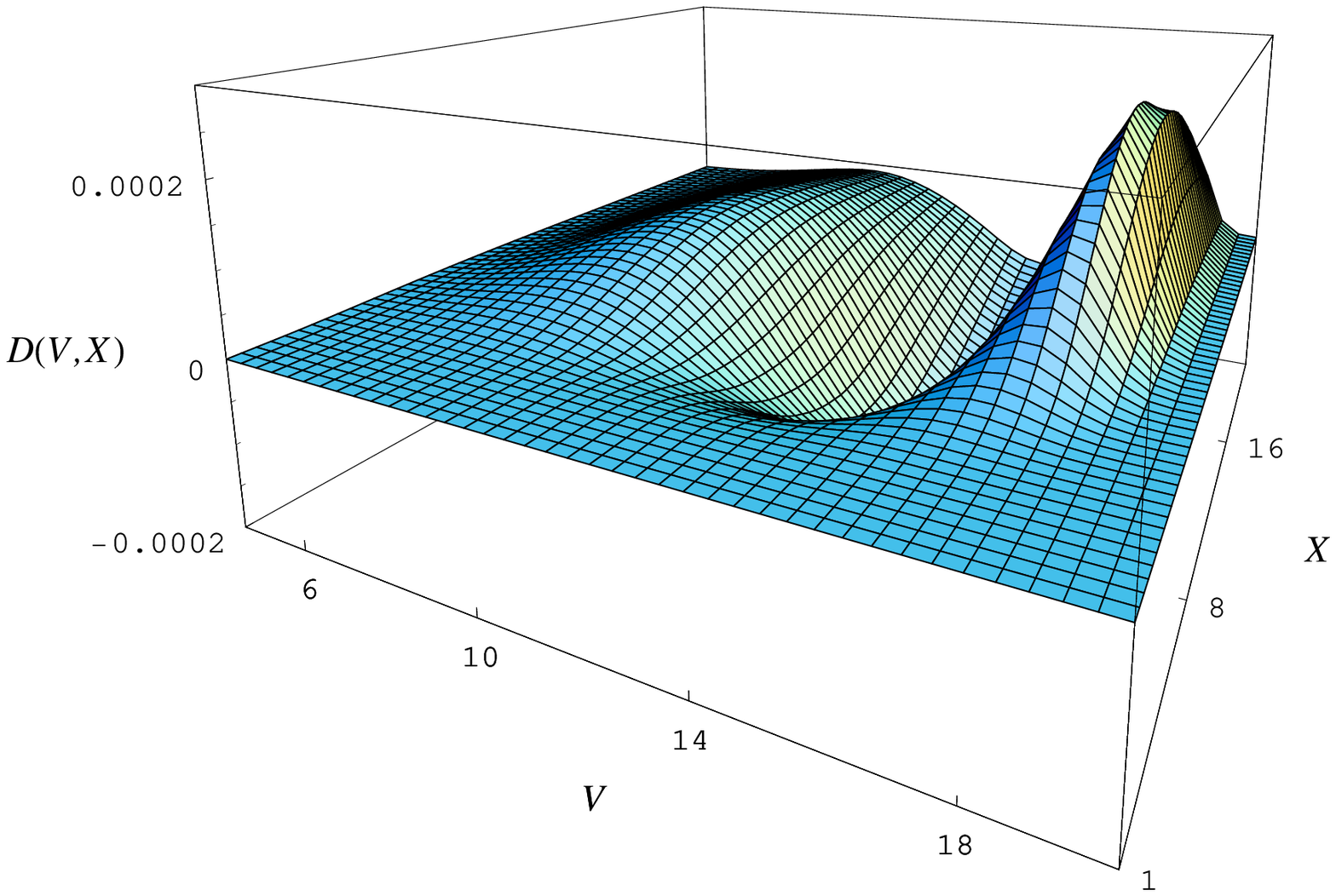}
\end{center}
\caption
{Difference of probability densities (7)   of  correlated and uncorrelated 
group of 5 cars for  the same parameters as in figure 5.
} 
\label{fig:11}
\end{figure}

All the peaks  corresponding to free and congested flow or formation of 
platoons result from correlations between the cars. It is seen, e.g., in 
Fig.~11, where the difference between the probability density $D(V,X)$ of 
correlated and uncorrelated group of 5 cars, which $P(V,X) $ is in Fig.~5, 
is shown. The    probability 
density of a group of uncorrelated cars always displays only a single peak 
without any internal structure.

\section{Discussion}

Deriving canonical distribution of an equilibrium  gas of classical particles 
in statistical mechanics,
all the 
velocities and positions of the particles are assumed to be equally probable, 
and a condition of given total mean energy is imposed on the system. (In
grandcanonical distribution a condition for number of particles is added.)   
In our 
approach to the gas of cars all their relative positions are equally probable 
as well, 
but conditional probabilities of their velocities are given by behaviour 
of drivers. Instead of the condition on total energy,  equality condition 
of mean velocities of all cars and a condition on the mean total length of the 
system are imposed. Now, the probability of a state has not a simple form of 
an exponential of the Hamiltonian like in the case of physical systems. In 
principle, it can be rewritten into such a form, but now the quantities 
appearing 
in the exponential are not conserved in an isolated system, what is the case 
in a gas of classical particles in physics.  That is why it is not possible to 
introduce an effective energy or Hamiltonian in our system of cars, i.e., our 
approach differs from that of thermodynamical traffic gas \cite{l:10} where 
interactions between cars and Hamiltonian of the system is introduced. 
Nevertheless, the results are similar but the velocity distribution of cars is 
not Gaussian.

Our method is, to some extent, similar  to a mean-field and cluster 
approximation \cite{l:6,l:7,l:8}  looking for  analytical solution of master 
equation 
corresponding to Nagel-Schreckenberg probabilistic cellular automaton model. 
Our choice of variables, velocities and headways,  corresponds to that in 
the car 
oriented mean field theory \cite{l:9}, where the central role plays the  
probability 
$D_n(v)$ identical to our $P(v_{i-1},x_{i-1,i})$.  In the above-mentioned 
approaches an approximate solution of master equations were found. In the 
present paper the solution is obtained by exact minimization of the entropy of 
the system.
 
Our results are consistent with a number of other approaches, based on kinetic 
or dynamical decripton of cars, and experimental observations of single line 
traffic \cite{l:4,l:11,l:12}. It seems that  systems of cars behave like 
many-particle physical systems or small parts of them, which, on microsopic 
level, are ruled by Newton or kinetic equations and, in equilibrium, 
 are practically always 
described by canonical distribution derived from the principle of maximum 
entropy.

\ack 
 We acknowledge support from VEGA grant No. 2/0113/2009.


\newpage


\end{document}